# Does the Position of Surgical Service Providers in Intra-Operative Networks Matter? Analyzing the Impact of Influencing Factors on Patients' Outcome

**Authors and affiliations:** Ashkan Ebadi, Postdoctoral Associate[1], Patrick J. Tighe, Associate Professor[2], Lei Zhang, Data Engineer[2], and Parisa Rashidi, Assistant Professor[1]

[1]Department of Biomedical Engineering, University of Florida, Gainesville, FL, USA
[2]Department of Anesthesiology, University of Florida, Gainesville, FL, USA

**Corresponding author:** Ashkan Ebadi, ashkan.ebadi@gmail.com

## Abstract

**Objective:** We analyzed the relation between surgical service providers' network structure and surgical team size with patient outcome during the operation. We also did correlation analysis to evaluate the associations among the network structure measures in the intra-operative networks of surgical service providers.

**Materials and Methods:** We focused on intra-operative networks of surgical service providers, in a quaternary-care academic medical center, using retrospective Electronic Medical Record (EMR) data. We used de-identified intra-operative data for adult patients, ages $\geq$ 21, who received nonambulatory/nonobstetric surgery in a main operating room at Shands at the University of Florida between June 1, 2011 and November 1, 2014. The intra-operative dataset contained 30,211 unique surgical cases. To perform the analysis, we created the networks of surgical service providers and calculated several network structure measures at both team and individual levels. We considered number of patients' complications as the target variable and assessed its interrelations with the calculated network measures along with other influencing factors (*e.g.* surgical team size, type of surgery).

**Results:** Our results confirm the significant role of interactions among surgical providers on patient outcome. In addition, we observed that highly central providers at the global network level are more likely to be associated with a lower number of surgical complications, while locally important providers might be associated with higher number of complications. We also found a positive relation between age of patients and number of complications.

**Conclusion:** Improving surgical quality is an important issue that requires a precise evaluation of all the potential influencing factors. Although we cannot infer causality, our findings suggest that brokers and leaders, *i.e.* providers with high betweenness and eigenvector centralities respectively, are associated, in a very general sense, with lower number of complications. On the other hand, we found that being close to many other surgical providers may not be advantageous for patients' outcome, in terms of number of complications. For certain types of surgeries, team consistency and surgical volume are associated with improved postoperative outcomes. However, each of these findings highlights the analytical potency of social network analysis in healthcare setting and points to the need for greater investigation linking medical domain specific mechanisms to the network structure measures.

**Introduction**

Surgical service providers are constantly interacting with each other as part of various teams to provide care to patients. While teams are generally comprised of members in well-defined roles, there is wide latitude in the composition of both the involved roles as well as the characteristics of individuals subsuming those roles. Although the surgical technology and procedure have improved, team modelling and role assignment are not yet well-established, resulting in miscommunication and mal-coordination among the team members [1]. That is, different team compositions might affect the overall performance of the team, due to communication breakdowns, mismatch of expertise, and other human factors. Although the benefits of an optimal team composition are obvious, forming a balanced team is difficult [2, 3]. That is, if the team is not balanced, conflicts and miscommunication among the team members might lead to overall weak performance. Thus, formation of teams in clinical care settings is an important problem that needs to be solved. Apart from the need for a model that captures structures and processes, and considers resource limitations, clinicians are required to evaluate the performance of the model/method periodically [1].

Although the definition of a team may vary across different domains, workplace teams, in general, have several common characteristics, such as: 1) being composed of two or more members who, 2) work on relevant tasks to reach a common goal, 3) have some independencies, *e.g.* knowledge level, 4) are embedded in an organizational context, and 5) have some organizational boundaries [4-6]. Several factors can influence the productivity of a team and/or improve communication. For example, there is a positive relation between the size of the team and intra-team process complexity [7]. Therefore, there is a trade-off between being a member of a larger team and the team's overall performance. Inherently, most individuals tend to collaborate with their past collaborators with whom they have had a successful experience [8]. This might cause a biased and unbalanced collaboration system in which some members are extremely active and collaborate in dense groups, while others work in sparse groups.

Time is also a crucial factor of team development [9], making the team formation a dynamic phenomenon [7]. That is, teams are formed, developed, and evolved over time [10], and the timing of this evolution may influence the team's performance. Team level characteristics emerge from the individual characteristics over time [6] such that behavior and characteristics of the team members and their interactions are reflected in the team. And, different teams might also interact in turn on a larger multilevel system [7] and evolve over time. Therefore, it is essential to consider the time effect in order to investigate team effectiveness [11], and individual and team level measures also need to be considered with respect to time to analyze the teams as well as their performance more accurately.

Improving surgical quality is an important issue that requires a precise understanding of the structures and processes that affect the surgical care. In addition to the pathophysiological risk factors, the surgery outcome also depends on the events and the quality of care that the patient receives during his/her stay in hospital [12]. This includes the performance of the surgical service providers as well as health professionals, both at team and individual levels. In a qualitative study, Main *et al.* [13] assessed the structures and processes of surgical care and their impact on quality and outcome of surgery, through interviewing surgical care providers and leaders from six hospitals. Their findings show that communication and care coordination are highly essential for an effective surgical service [13]. This highlights the crucial role of providers' interactions in surgical outcome.

Let us review two sample scenarios in our examined data to further illustrate the potential and crucial role of providers' interaction on patients' outcome. We selected two random patients with similar properties from our dataset, called patients A and B in Fig. 1, who went under the same surgery. As seen in Fig. 1, the surgical team size and composition for patients A and B are also comparable, however, the outcome of the surgeries differs significantly in terms of the number of postoperative complications. Although several factors could cause such difference in complications, from the figure, it is observed that the network structure property of the surgical team members who performed the operation on patients A and B are considerably different. More specifically, in the middle panel of Fig 1, the circle sizes reflect the betweenness centrality[i] of providers such that higher betweenness is shown with larger circles. Patient B was operated with a surgical team whose members were more central, resulting in zero postoperative complication. The story for patient A is reverse. Needless to mention that this observation does not prove anything, but perceiving similar observations in different random checks was our main motivation to perform this study.

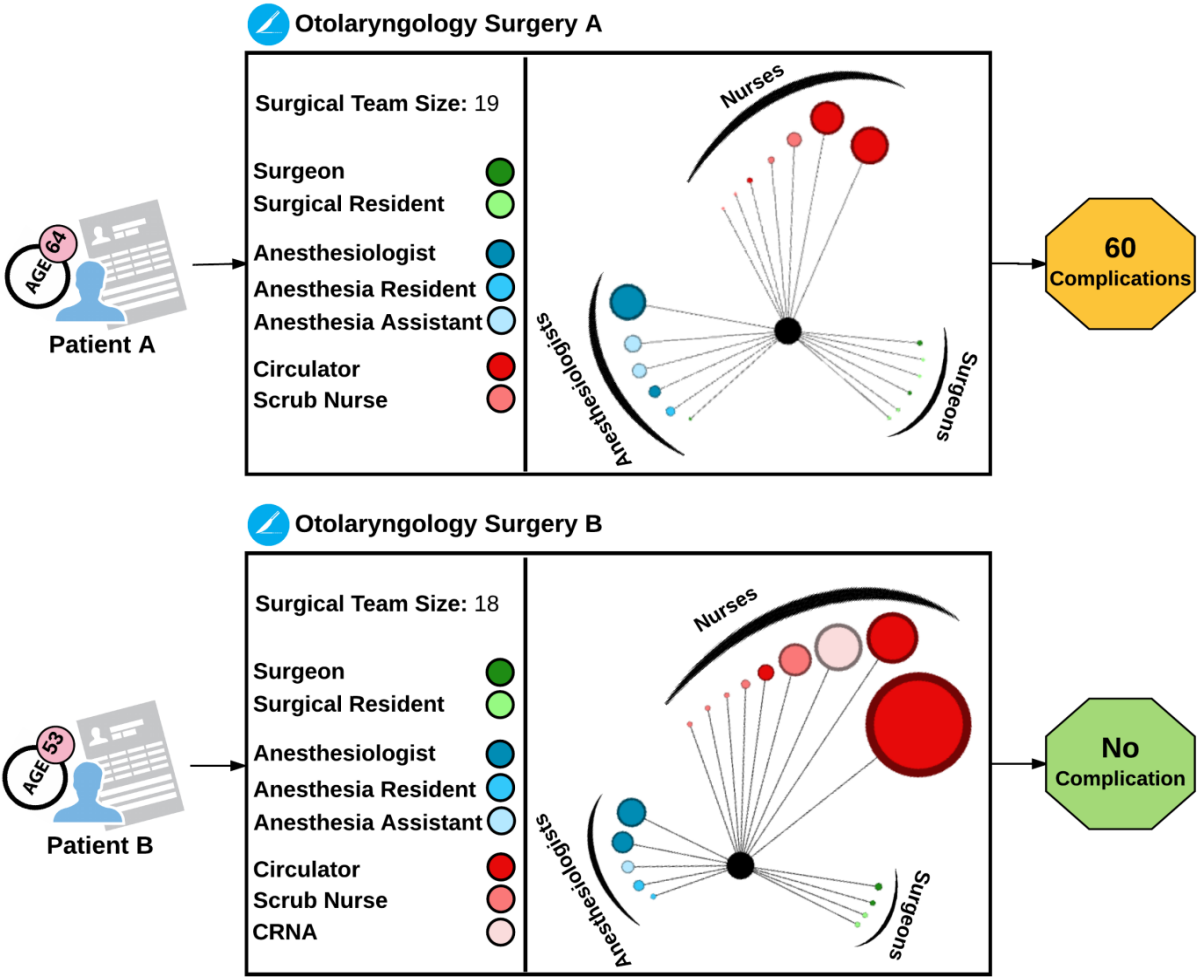

**Fig 1.** A sample scenario of two similar patients underwent the same surgery but resulted in different outcomes. As seen, network structure properties of the surgical service providers is one of the important potential factors that might have contributed to different outcomes for the given patients. The circle sizes in the right side of the middle panel reflect the betweenness centrality of the providers.

In this paper, we used both team and individual level features to characterize surgical teams and to investigate the relationship of surgical team structure on patient outcome in terms of the number of complications following surgery. In particular, we focused on intra-operative networks of surgical service providers, in a quaternary-care academic medical center, using retrospective Electronic Medical Record (EMR) data. The number of patients' complications was considered as a performance proxy for the surgical teams. To the best of our knowledge, this is the first paper that focuses on perioperative teams at the enterprise scale to analyze the inter-relations between various network measures and patients' outcomes, using both team and individual levels features.

---

[i] For the definition, please refer to the Data and Methodology section.

**Data and Methodology**

*Data*

This study was approved by the University of Florida Institutional Review Board. The data were collected from University of Florida's Integrated Data Repository (IDR) after obtaining a confidentiality agreement from the IDR. It contained de-identified intra-operative data for adult patients, age greater than or equal to 21, who received non-ambulatory/non-obstetric surgery in a main operating room at Shands at the University of Florida between June 1, 2011 and November 1, 2014. The de-identified surgery dates indicated the number of days elapsed from a common undisclosed original date prior to the study period. Subjects who did not receive a surgical procedure, or who were discharged on the same day as their surgery, were excluded from the dataset. Those cases whose de-identified surgery start and/or end dates were missing were also excluded. We also removed any invalid, generic or placeholder provider entries, *e.g.* providers whose ids were missing. The intra-operative dataset, named as the *Original Data* in Fig , contained 30,211 unique surgical cases.

The de-identified time order of the medical cases was used to slice the data into four separate, sequential time intervals each containing 365 temporally-contiguous events (one year). We will refer to each time slice as a *segment* in the rest of the paper, numbered from *segment 1* to *segment 4*, respectively in Fig . This step was necessary to differentiate between different co-worker networks as networks might evolve and change from time to time. We chose a one-year time window for creating the surgical providers co-worker networks, as opposed to computing the network measure for the entire network over four years, since providers might join the system or leave in the course of one year. Moreover, trainees are generally appointed to one-year intervals, although start dates may stagger throughout the year. The healthcare system in question generally experienced at least 10% growth in global surgical volume each year. Thus, the network might change or evolve each year. In addition, creating a single network for the entire dataset makes it impossible to observe the temporal effects. After creating the four segments, we calculated several network structure measures for each of the providers in each time frame. We computed the average network measures of surgical team members to obtain the network measure for each intra-operative surgical team. For this purpose, teams of providers were retrieved for each surgical case, de-identified providers were listed, their network measures were summed up for each measure separately, and measures were averaged over the size of the team.

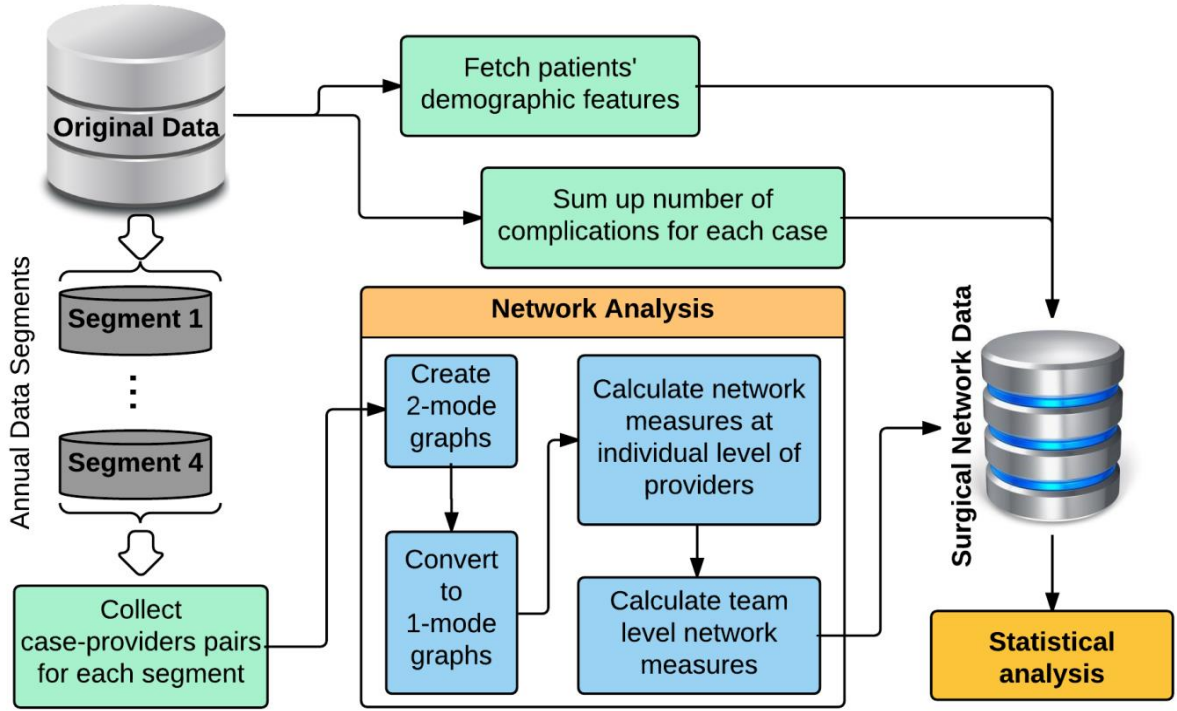

**Fig 2.** The data collection and methodology in brief. The original intra-operative data was first sliced into four separate, sequential time segments, each representing one year of data. For each segment, all the <case, providers> were collected and provided to the network analysis module. In the network analysis module, the networks for each of the four time slices of the intra-operative dataset were constructed and network measures were calculated for each network. The calculated measures along with patients' demographic data and total number of complications for each medical case were integrated into a single dataset named, the Surgical Network Data, which was used for performing the analysis.

A list of patient socio-demographics is given in Table 1. Providers were considered from intraoperative phases of care. In the intraoperative phase, providers were those individuals who were documented as having participated in a patient's surgery. The majority of these roles were surgeons and surgical assistants, anesthesiologists and anesthetists, and circulating nurses. As a teaching hospital where trainees may rotate to different services, trainees were included using their denoted roles within the given surgery.

**Table 1.** List of variables, their types, and description

| Variable | Type | Description |
|---|---|---|
| *Age* | Numeric | Patient age at hospital encounter (set as 90, if age 90 or older) |
| *Race* | Character | Patient's race |
| *Ethnicity* | Character | Patient's ethnicity |
| *Gender* | Character | Patient's gender |
| *Marital Status* | Character | Patient's marital status at hospital encounter |
| *BMI* | Numeric | Patient's Body Mass Index |
| *Comorbidity* | Numeric | Patient's Charlson Comorbidity Index (ver. 2011) |
| *LOS* | Numeric | Length of Stay (Inpatient/Observation in days or in hours) |
| *Service* | Character | Type of surgical service |
| *Principal Dx* | Character | Primary diagnosis code and description |
| *Principal Px* | Character | Primary procedure code and description |
| *Complication count* | Numeric | Derived from diagnosis1-diagnosis50 and complication codeset |

Surgical complication codes were defined as the codeset of a list of ICD9-CM codes, denoting various surgical complications. The complication ICD9-CM codeset is listed in Appendix A. The diagnoses are coded during an abstraction of the medical record which occurs at the conclusion of the hospitalization. The ICD9-CM complication codeset contains codes from 996 through 999, along with their corresponding subclasses. We scanned through each of 50 diagnosis codes to check if any of the complication codes matches. The complication outcome variable, which is used in statistical analyses, is the sum of all detected complication codes for a hospital encounter. Surgical complications do not necessarily denote any error in the surgical process, as several complications are outcomes which may be unavoidable given the nature of the patient's comorbidity status and/or the nature of the procedure itself.

*Network Analysis*

In a network, nodes are connected to each other through edges. Surgical service providers can be considered as nodes in intra-operative network, and any type of relationship between them, such as being involved in a surgery, can be regarded as edges that connect the network nodes to each other. The manner in which the nodes are connected to each other varies in different networks, thus networks exhibit diverse characteristics which can be measured by different network measures. A network is called an *undirected network* if all the edges are bidirectional, and a *directed network* if edges are directed from one node to another.

As mentioned earlier, we built surgical service providers' networks for each of the four time segments of the intra-operative dataset. These networks are referred to as *network 1* to *network 4* in the rest of the paper. To create the networks, we first generated the two-mode networks [14]. Two-mode networks are called *bipartite graphs* in graph theory in which nodes can be divided into two disjoint sets, and edges can only connect a node from one set to a node in another set, hence there is no inside-set connection. In our two-mode networks, surgical service providers, *e.g.* surgeons, anesthesiologists, and circulating nurses, were connected through medical cases serviced by them. In other words, in our two-mode networks all the medical service providers who were involved in a medical case are connected to the case (Fig , left). Since service providers were not connected to each other in the two-mode networks, we converted each of the created two-mode networks to one-mode networks in which surgical service providers are connected to each other if they have been working together on the same case (Fig , right).

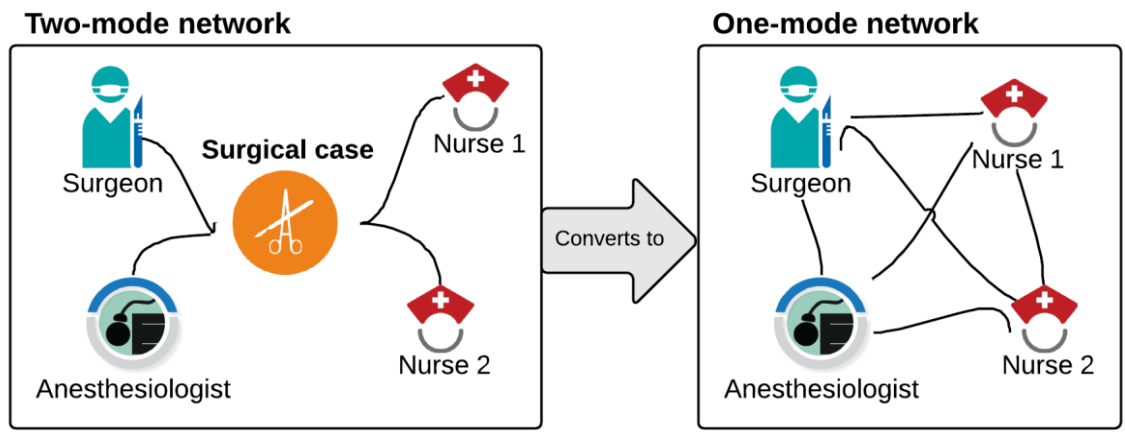

**Fig 3.** Two-mode vs. one-mode networks for a hypothetical surgical case. The two-mode network is converted to a homologous one-mode structure by removing the case from the two-mode networks. The edges in the one-mode network thus denote a connection between providers via a shared surgical case.

After creating the one-mode networks for each of the four data segments, we next calculated several network structure measures for each node, *i.e.* for each surgical service provider in each of the networks. We selected several commonly used network measures, *i.e.* betweenness centrality, closeness centrality, degree centrality, clustering coefficient, and eigenvector centrality, for statistical modeling and correlation analysis. Three network variables, *i.e.* eigenvector, closeness, and betweenness centralities were used in the regression analysis, while we considered all the five network measures in the correlation analysis. We were forced to exclude two network measures from the regression analysis due to the multicollinearity problem that is explained later. The network measures are briefly introduced in this section.

*Degree Centrality*: Degree centrality is a simple network structure measure which is defined based on the number of edges incident to a node, *i.e.* the degree [15]. Degree centrality can be regarded as a proxy for the number of direct connections that each node has in a network. In the case of surgical service providers, higher degree centrality can indicate higher importance of the node. That is, higher number of direct connections, and being involved in more surgeries might bring an advantage to the provider and increase his/her influence, through being connected with/knowing more other providers.

*Betweenness Centrality*: Betweenness centrality is one of the essential graph-theoretic concepts and a standard measure of control in network analysis [16]. This measure focuses on an important concept in networks, *i.e.* the *brokerage* role, which is the ability of the nodes to bridge different group of nodes, and/or control communities in a network [17]. In our case, betweenness centrality measures the extent that a surgical service provider is positioned on the shortest path between any other pairs of providers in the intra-operative network, thus, bringing them a strategic position/advantage in the network, even at the global level. Being on the path of information pool as well as having access to a variety of clusters/communities, would theoretically enable brokers to be well aware of the information flow, to find new/better partners, or to get involved in new/better performing teams. Mathematically, betweenness centrality of node $i$ ($bc_i$) is defined based on the share of times that a node $j$ reaches a node $k$ via the shortest path passing from node $i$ [18-20], as stated in Equation (1).

$$bc_i = \sum_{i \neq k \neq j} \frac{\sigma_{jk}(i)}{\sigma_{jk}}. \qquad (1)$$

In Equation (1), $\sigma_{jk}$ is the total number of shortest paths from node $j$ to $k$, and $\sigma_{jk}(i)$ is the number of shortest paths from node $j$ to node $k$ that pass through node $i$.

*Eigenvector Centrality*: Bonacich [21] in 1972 stated that the eigenvector of the largest eigenvalue of a connectivity matrix can be used as a good network centrality measure. Eigenvector centrality is argued to have some advantages over conventional graph-theoretic measures, such as degree centrality [22]. In degree, each contact is weighted equally, however, in eigenvector contacts are weighted based on their centralities. That is, apart from the node itself, eigenvector centrality also considers the connections of the node. A node with high eigenvector centrality is connected to other nodes which occupy important positions in the network. Thus, the eigenvector centrality can be regarded as a weighted sum of not only direct connections, but indirect connections of every length, which enables it to better capture the entire pattern in the network [22]. In our case, it can be seen as a proxy of a surgical service provider's influence at the global level of the network. Being connected to other highly influencing providers can bring a strategic and diplomatic power to a surgical provider, which makes the role interesting for our analysis. Bonacich [21] defined the eigenvector centrality of node *i* ($c_i$) based on the sum of the centralities of its adjacent nodes, as in Equation (2).

$$c_i = \lambda \sum_{\{i,j\} \epsilon E} c_j. \qquad (2)$$

In equation (2), *i* and *j* are two given nodes in a graph that are connected by the edge *{i,j}* and $\lambda$ is the normalization factor. Surgical providers with high eigenvector centrality in intra-operative networks might play a *leadership* role, because they are connected with too many other influential and highly central providers. It is hence expected that they shape the interactions and play an important role in setting priorities.

*Closeness Centrality*: Closeness centrality is a key node centrality measure in network analysis [15,23]. This measure is defined based on the shortest path between nodes in a network, and considers both direct and indirect connections among the nodes [24]. Closeness centrality indicates how close a node is to the other nodes in a network. Thus, nodes with high closeness centrality can be considered as *local influencers*, as they are not only highly connected to other nodes, but relatively close to them, at least at their own community level. Local influencers are able to facilitate the interactions and communication with other nodes in a network, if required, hence bringing an advantage to themselves, as well as to their surrounding community. One should note that local influencers are not necessarily important at the global network level but, they are often locally very important ones, as they can have an influence on the information spread, and on the access to the key resources, *e.g.* human resource. Closeness centrality of node *i* is defined as in Equation (3).

$$cl_i = 1 / \left( \sum_{j \in n-\{i\}} d(i,j) \right). \qquad (3)$$

In Equation (3), *n* is the number of nodes in the network, and *d(i,j)* is the length of the shortest path between the nodes *i* and *j*. From Equation (3), it is clear that closeness centrality can be only calculated in the connected components (a sub-network in which there is no isolated node and all the nodes are interconnected) or networks. Because the distance is infinite between nodes in disconnected components, the denominator becomes ∞, and as a result the closeness centrality will be zero, which is not informative [25-28]. But, this limitation is justifiable as the core interactions mainly occur in the largest component [29], especially if the size of the largest component is significant.

*Clustering Coefficient*: Clustering coefficient represents the tendency of the nodes in a network to cluster together, and it shows how well-connected the vicinity of the node is. Nodes with high clustering coefficient are more likely to create dense inter-connections around them, thus, creating tightly knit groups [30]. In the case of intra-operative networks, clustering coefficient can quantify how close a provider and his/her co-workers are to become a complete sub-network. Theoretically, clustering coefficient of node *i* ($CC_i$) in network *G* is defined as stated in Equation (4).

$$CC_i = \frac{\lambda_i^G}{\tau_i^G}. \qquad (4)$$

In Equation (4), $\lambda_i^G$ is the number of subnetworks in network *G* with three edges and three nodes, $\tau_i^G$ is the number of triples on node *i*, *i.e.* the number of subnetworks with three nodes, including *i*, and two edges such that node *i* is connected to both other two nodes.

*Statistical Analysis*

As the first step, we did a correlation analysis on all the five calculated network structure measures, *i.e.* betweenness centrality, degree centrality, closeness centrality, eigenvector centrality, and clustering coefficient. Apart from analyzing the inter-relations among the network structure measures in the examined intra-operative dataset, this step was also helpful in selecting the final list of independent variables to be used in the second part of the statistical analysis. In particular, we performed Spearman correlation analysis [31], because we suspected a monotonic relationship between pairs of network measures. In addition, bivariate non-normal distribution was also expected among the network measures. This step also included summative descriptions on the size of the network and its components.

As the second part of the statistical analysis, we selected the number of complications in each case as the dependent variable, since the purpose of this study was to analyze the impact of the influencing factors and network measures on patient outcome. Our dependent variable is therefore a count measure. The Poisson model is normally proposed for count dependent variables [32]. Although the best matching regression model is Poisson, in reality it is rare to satisfy the Poisson assumption on the actual distribution of a natural phenomenon, because most of the time an over-dispersion or under-dispersion is detected in the sample data. This causes the Poisson model to underestimate or overestimate the standard errors and thus results in misleading estimates for the statistical significance of variables [33]. We empirically tested Poisson model on our data and found that Poisson model does not fit to our data because the goodness of fit chi-squared test was statistically significant. In particular, we obtained a large value of 313833.6 for chi-square which is an indicator that the Poisson model is not fitting to the data very well. In addition, a significant p-value of (0.000) was observed for the goodness of fit statistic, confirming that Poisson is not appropriate. In order to obtain robust standard errors and to correct the estimates, negative binomial regression can be employed as an alternative to Poisson regression [32]. Thus, we employed negative binomial regression on our data to estimate the impact of the selected factors on patients' number of complications.

In negative binomial regression, which is a generalized linear model, the dependent variable is a count of number of times that an event occurs. Equation (5) represents the negative binomial distribution [34].

$$p(Y = y) = \frac{\Gamma(y + \frac{1}{\alpha})}{\Gamma(y+1)\Gamma(\frac{1}{\alpha})} \left(\frac{1}{1+\alpha\mu}\right)^{\frac{1}{\alpha}} \left(\frac{\alpha\mu}{1+\alpha\mu}\right)^y. \qquad (5)$$

In Equation (5), $Y$ is the dependent variable, $\mu > 0$ is the mean value of Y, and $\alpha > 0$ is the heterogeneity parameter. Based on [34] definition, $\frac{1}{\alpha}$ should not be necessarily an integer. From Equation (5), Hilbe [34] defines the negative binomial regression model as stated in Equation (6).

$$\ln \mu = \beta_0 + \beta_1 x_1 + \beta_2 x_2 + \cdots + \beta_p x_p. \qquad (6)$$

In Equation (6), $x_1, x_2, \ldots, x_p$ are the given independent variables, and $\beta_0, \beta_1, \ldots, \beta_p$ are the regression coefficients which are needed to be estimated. Having selected the negative binomial regression as the model, we prepared a primary list of exploratory candidates to include them in the model. Next, as suggested in [35], we checked for various meaningful combinations of the listed independent variables, added them to the model, and tested the results to obtain the final list of the independent variables which produces the most significant and robust results.

The correlation among the candidates was also checked. The reduced form of the model is stated in Equation (7).

$$C_i = f(age_i, teamSize_i, typSurgery, avgBtwn_i, avgClos_i, avgEigen_i, dMale_i). \quad (7)$$

The variables description is listed in Table 2. Including patients' age and gender partially adjust the impact of the network structure measures. The surgery type variable controls for the fact that the complication patterns might vary in different surgeries. We used STATA 12 data analysis and statistical software to perform the statistical analysis.

**Table 2.** List of variables, their types, and description

| Variable | Type | Description |
|---|---|---|
| $C_i$ | Dependent | Number of complications occurred in the i[th] medical case. |
| $age_i$ | Independent | Age of the patient in the i[th] medical case. |
| $teamSize_i$ | Independent | Number of surgical service providers in the i[th] medical case. |
| $typSurgery_i$ | Independent | Type of the surgery in the i[th] medical case. |
| $avgBtwn_i$ | Independent | Average betweenness centrality of the surgical providers in the i[th] medical case. |
| $avgClos_i$ | Independent | Average closeness centrality of the surgical providers in the i[th] medical case. |
| $avgEigen_i$ | Independent | Average eigenvector centrality of the surgical providers in the i[th] medical case. |
| $dMale_i$ | dummy | Equals to 1 if patient in the i[th] medical case was male, otherwise 0. |

**Results**

**Descriptive Analysis**

The non-segmented intra-operative network contains 30,211 surgical cases in total, involving 1,682 distinct surgical providers (nodes), and 203,224 connections among providers (edges). The entire network is highly connected with the average degree of 241.6. The average surgical team size for the entire network is 8. Table 3 shows the data distribution, in terms of the number of nodes and edges, in different networks of the intra-operative dataset. Moreover, the network structure measures maintain similar levels in different segments. As seen, the networks are generally comparable, in terms of number of nodes and edges, except for the number of edges in the last network which was expected, since the time window for the final segment is slightly smaller in comparison with the other segments. Thus, the resulting networks and the respective findings for each segment can be compared to one another. The density measure is calculated by counting the number of edges in a given graph and dividing it by the maximum number of possible edges between the nodes in the given graph. Although the networks corresponding to each data segment in the intra-operative dataset are not extremely dense according to the density measure, the average degree is relatively high indicating a relatively large average number of direct partners for each team member.

**Table 3.** Data distribution in different segments of intra-operative dataset

|  | Network | | | |
|---|---|---|---|---|
|  | 1 | 2 | 3 | 4 |
| **Nodes** | 896 | 914 | 978 | 761 |
| **Edges** | 74,632 | 78,789 | 84,095 | 42,377 |
| **Cases** | 8,226 | 8,541 | 9,257 | 4,187 |
| **Average Team Size** | 8.1 | 8.2 | 8.3 | 6.8 |
| **Average Degree** | 166.6 | 172.4 | 172 | 111.4 |
| **Density** | 0.186 | 0.189 | 0.176 | 0.147 |
| **Average Betweenness** | 0.002 | 0.002 | 0.002 | 0.003 |
| **Average Closeness** | 0.59 | 0.59 | 0.58 | 0.57 |
| **Average Eigenvector** | 0.58 | 0.53 | 0.58 | 0.54 |
| **Average No of Complications** | 2.2 | 2.3 | 2.1 | 1.3 |

**Network Measures, Correlation Analysis**

As the next step, we did correlation analysis to evaluate the relations and linear associations among the network structure measures in the intra-operative networks of surgical service providers. For this purpose, we chose Spearman correlation because a monotonic relationship between any pairs of the network structure measures was expected, and the fact that bivariate normal distribution assumption is not required in Spearman correlation. Table 4 shows the Spearman's correlation matrix of the intra-operative network structure measures. As observed, all correlations were significant at the confidence level of 99%. A strong monotonic and positive relationship ($\geq 0.8$) was seen between three pairs: 1) closeness centrality and eigenvector centrality, 2) closeness centrality and degree centrality, and 3) eigenvector centrality and degree centrality. This was expected, since a node with higher degree is more likely to be connected to other important nodes in the network, or to be on the shortest paths among the other nodes. The negative relationship between clustering coefficient and betweenness centrality points to the fact that in surgical service providers networks, gatekeepers are not themselves highly clustered. Gatekeepers can be regarded as providers with high betweenness centrality who have a control over their surrounding network and the flow of information. Thus, although gatekeepers can connect distinct and different clusters in the network, their immediate surrounding group/community is not tightly clustered. The same argument is valid for the relationship between the eigenvector centrality and the clustering coefficient. Notably, the negative relationship between eigenvector centrality and the clustering coefficient highlights an important property in the network of surgical providers. That is, the influencing providers who are connected to a provider with high eigenvector centrality, let us say the *core*, are not very likely to be connected to each other, thus mostly relying on the core to link them to other influential providers. This points out the leadership/supervisory role of the provider with high eigenvector centrality in the examined network.

**Table 4.** Spearman's correlation matrix of network structure measures

| **Variable** | *avgBtwn* | *avgClos* | *avgEigen* | *avgClust* | *avgDeg* |
|---|---|---|---|---|---|
| *avgBtwn* | 1.00 | | | | |
| *avgClos* | 0.53* | 1.00 | | | |
| *avgEigen* | 0.54* | 0.84* | 1.00 | | |
| *avgClust* | -0.69* | -0.3* | -0.45* | 1.00 | |
| *avgDeg* | 0.37* | 0.91* | 0.8* | -0.2* | 1.00 |

Note: * $p<0.01$, number of observations: 30,211

**Regression Analysis**

In this section, we discuss the negative binomial regression results. The number of complications (*C*) is the dependent variable, and a set of network structure measures, *i.e.* average betweenness centrality (*avgBtwn*), average closeness centrality (*avgClos*), and average eigenvector centrality (*avgEigen*), along with age (*age*), type of surgery (*typSurgery*), and gender of the patient (dMale), and number of providers in surgical teams (*teamSize*), form the set of independent variables. We first listed all the meaningful variables as independent variable candidates. This list included but not limited to patient's race and ethnicity, Charlson morbidity index and type of surgery, and a comprehensive list of network structure variables. Then, we performed several diagnostic tests to identify the final list of predictors. All possible combination subsets of the predictors were created and tested, and the set of predictors with the most robust results was finally selected. The correlation analysis was necessary as a preliminary step. We also checked for the multicollinearity among the candidate predictors by calculating the variance inflation factor (VIF). VIF quantifies the severity of multicollinearity. We performed an ordinary least squares regression analysis on all the listed variables, and calculated VIF for the predictors. Using the observed associations between the network variables, we selected the final list of network variables to include in the regression model. We observed 1/VIF to be less than 0.1 for all the predictors, indicating that the degree of collinearity is not significant.

As seen in Table 4, the absolute value of all the correlation coefficients is lower between closeness centrality, betweenness centrality, and eigenvector centrality. Thus, these three variables were included in the regression model, representing the network structure. We did not include the average degree as we already had the team size in our model, which is highly correlated with the average degree. Table 5 shows the results of the regression analysis.

**Table 5.** Negative binomial regression results. Number of complications (*C*) is the dependent variable. Patient's age, surgical team size, type of surgery, average betweenness, closeness, and eigenvector centralities are the independent variables.

| *C* | *Coefficient* | *Std. Error* | *Z* | *P>|z|* | *[95% Confidence Interval]* | |
|---|---|---|---|---|---|---|
| *age* | 0.007*** | 0.001 | 4.53 | 0.000 | 0.004 | 0.009 |
| *teamSize* | 0.154*** | 0.011 | 14.59 | 0.000 | 0.134 | 0.175 |
| *typSurgery* | 0.017*** | 0.002 | 8,59 | 0.000 | 0.013 | 0.021 |
| *avgBtwn* | -62.258* | 33.554 | -1.86 | 0.064 | -128.023 | 3.506 |
| *avgClos* | 11.2*** | 2.73 | 4. 1 | 0.000 | 5.85 | 16.55 |
| *avgEigen* | -1.856*** | 0.578 | -3.21 | 0.001 | -2.99 | -0.722 |
| **Gender dummy variable** | | | | | | |
| *dMale* | 0.1** | 0.046 | 2.16 | 0.031 | 0.009 | 0.191 |
| *_cons* | -6.859*** | 1.321 | -5.19 | 0.000 | -9.447 | -4.271 |
| *ln(alpha)* | 2.732 | 0.016 | | | 2.7 | 2.763 |
| *alpha* | 15.357 | 0.249 | | | 14.877 | 15.853 |
| Likelihood-ratio test of alpha=0 | | | chibar2(01) = 1.6e+05 | | Prob>=chibar2 = 0.000 | |

Note: * p<0.10, ** p<0.05, *** p<0.01, number of observations: 30,211

According to Table 5, all the independent variables significantly impact the number of complications. Notably, the significance of the network measures is much higher than the other variables, indicating the high importance and significant effect of inter-relations in intra-operative segment on patient outcome. Although, as expected, there is a positive relationship between age of the patient and number of complications, the coefficient is very small (0.007). This is in line with the literature, where although it is agreed that elder patients are more likely to experience adverse surgical outcomes, the magnitude of such association is not very clear [36]. The positive impact of surgical team size on the number of complications is of great importance which partially indicates the significance of optimal team arrangement in healthcare settings. That is if the surgical team is large, it will increase the associated probability of experiencing intra-process complexities [7], and potentially miscommunication among the team members, which could lead to higher number of complications. Thus, our results confirm the trade-off between the surgical team size and performance of the team, in terms of number of complications. We also observed a small positive relation (0.017) between the surgery types and the number of complications.

From Table 5, the effect of network structure measures on the number of complications is higher than the other types of independent variables. Although observing larger coefficient for the network structure variables does not necessarily imply higher importance, their magnitude sheds some lights on the influence of network structure and interactions amongst the surgical service providers on patients' outcome. The significance is the highest for the betweenness centrality (-62.258), and the lowest for the eigenvector centrality (-1.856), both with a negative impact. Observing a negative effect of eigenvector and betweenness centralities along with a positive impact of closeness centrality (11.2) on number of complications, further confirm the importance of collaboration patterns. Providers with high closeness centrality can be identified as important local influencers within their local network of surgical service providers or community. However, betweenness centrality and eigenvector centrality are more global network level measures. Meanwhile, as we already discussed (Table 4) in our examined intra-operative network, surgical service providers with high betweenness or eigenvector centralities are not highly clustered. Thus, according to our results, being relatively close to so many other providers might increase the patients' risk. Yet, being highly central at the global level might lower down the chance of patients' complications. This is in line with studies in other domains that found higher centrality will lead to lower chance of failure in complex systems (*e.g.* [37]).

Finally, the analysis of the gender dummy variable (*dMale*) revealed that there is a significant difference between male and female patients in having complications during the surgery. As seen in Table 5, the positive coefficient of *dMale* indicates that the number of complications is expected to be higher for the male patients in comparison with females. This is also in line with several studies that found a positive relation between the gender and patient outcome such that, for example, female patients recover faster, or they need less intensive care in comparison with male patients (*e.g.* [38-39]).

**Conclusion and Discussion**

In this paper, we analyzed a large dataset of intra-operative interactions among surgical service providers. Our goal was to investigate the impact of various influencing factors of different types on patients' outcome. We considered both team level (*e.g.* team size) and individual level (*e.g.* network measures) variables, and statistically analyzed the inter-relations. We also accounted for the possible different impact of the surgery types on the number of complications. Our results confirm the significant role of network structure measures, and inter-actions amongst surgical service providers, in intra-operative patient care.

It was observed that brokers and leaders, *i.e.* providers with high betweenness and eigenvector centralities respectively, are associated, in a very general sense, with lower number of complications. The methods presented in this current manuscript are unable to infer causality for this interesting observation. Network brokers and leaders may be associated with lower numbers of complications due to their ability to access important skills, information, and consultations with other teams. Alternatively, brokers and leaders in this intra-operative network may reflect those individuals providing high-volume care for surgical cases that carry relatively low risk. Network leaders with high eigenvector centralities may reflect attending physicians who network with a range of highly-connected residents and nurses, thus reflecting their role as resources of experience. For the case of network brokers, betweenness centrality may point towards consultants who bridge the gap between disparate areas of sub-specialization; notably, high degrees of such consultancy may also point to more complex delivery of healthcare, and indeed would have suggested a higher rate of complications along this line of reasoning. Each of these possibilities points to the need for greater investigation linking healthcare delivery mechanisms to the observed network structures reported here.

In addition, an important fact can be inferred from the positive coefficient observed for the closeness centrality. That is, our results suggest that being close to many other surgical service providers in intra-operative segment may not be advantageous for patients' outcome, in terms of number of complications. This offers an intriguing counter-insight regarding the inverse associations between betweenness/eigenvector centrality and post-operative complications. These offsetting observations suggest that working, directly, with many others may be generally associated with adverse patient outcomes. Again, this may simply be a reflection of the fact that higher-complexity surgeries require larger teams, and are intrinsically carry more risk. Increasing closeness centrality may also be explained for procedures that have longer durations of surgery (and thus require more intraoperative breaks and handoffs for staff), which would both increase the closeness and reflect the increased risk for longer, and thus presumably more complex, surgeries compared with shorter, simpler and lower-risk procedures. For instance, a complex revision of an infected aortic graft in a heart-lung transplant recipient with renal failure is a riskier procedure which takes longer and requires more separate domains of expertise than an uncomplicated appendectomy for a healthy 22 year-old. In such cases the individuals would be expected to work directly alongside each other, rather than via indirect relationships via consultancy and expert-references. Finally, these findings are loosely aligned with data suggesting that for certain types of surgeries, team consistency and surgical volume are associated with improved postoperative outcomes.

The analysis of the inter-relations of age and gender of the patients with number of complications also shed some lights on the importance of personalized and patient-centered care system. Patients, as the most significant elements of health-care systems, should be treated with absolute attention. This is more critical for elder patients, as according to our findings, they are more open to higher risks and more complications in the perioperative setting, a finding that agrees with prior data. Meanwhile, the necessity of a gender-specific attention is also suggested as males are more likely to have more number of intra-operative complications according to our findings, which again is in keeping with prior reports. Notably, each of these findings was consistent with prior literature as well as after controlling for network effects, thus pointing towards the contribution of network effects to models of surgical outcome. Moreover, even coarse indices network effects carried significant weight in estimating the risk of a post-operative outcome, much more so than the acknowledged influence of age and sex.

**Limitations and Future Work**

Many of the limitations of this report represent the inherent constrains in working with retrospective electronic medical record data. As one of the first at-scale analyses of intra-operative networks, this approach examined aggregate findings of standard network measures with minimal stratifications. This approach was necessary to lay the groundwork for future investigations that can leverage these findings to consider the effects of detailed patient and healthcare provider characteristics, the dynamicity of network structure, the nature of interactions, separate phases of perioperative care delivery, weighting of healthcare provider and patient interactions across different dimensions of interest, and myriad related questions surrounding the structure of perioperative teams.

Perhaps the chief limitation of this study was the way we measured the connections among the surgical service providers. We were unable to catch the informal relations and interactions among providers, *e.g.* friendship, that might have an impact on the team arrangement procedure. This type of data is never recorded but definitely has an impact on the network structure. Manual collection of such data is resource-intensive and limits the scale of application. More automated measures, such as through the use of tracking technology, would permit the scaling of proximity capture, but carries notable concerns for both worker privacy as well as loss of detail on the context of temporospatial connectivity. EMR-based approaches complement other approaches to network data collection given the size and quality of the available data.

In summary, our findings point towards the importance of even coarsely-aggregated data on intraoperative healthcare provider network structures in influencing surgical outcomes. Substantial efforts are necessary to both further characterize perioperative network structures but also determine methods for incorporating these findings into more traditional statistical modeling approaches for predicting surgical outcomes.

**Acknowledgement**

This research was supported by the National Institute of General Medical Sciences (NIGMS) K23 GM102697 and NIGMS R01 GM114290.

**Appendix A**

**Table A. 1.** Complication ICD9-CM codeset.

| No | Code | Code Definition |
| --- | --- | --- |
| 1 | 996.0 | Mechanical complication of cardiac device, implant, and graft |
| 2 | 996.1 | Mechanical complication of other vascular device, implant, and graft |
| 3 | 996.2 | Mechanical complication of nervous system device, implant, and graft |
| 4 | 996.3 | Mechanical complication of genitourinary device, implant, and graft |
| 5 | 996.4 | Mechanical complication of internal orthopedic device, implant, and graft |
| 6 | 996.5 | Mechanical complication of other specified prosthetic device, implant, and graft |
| 7 | 996.6 | Infection and inflammatory reaction due to internal prosthetic device, implant, and graft |
| 8 | 996.7 | Other complications of internal (biological) (synthetic) prosthetic device, implant, and graft |
| 9 | 996.8 | Complications of transplanted organ |
| 10 | 996.9 | Complications of reattached extremity or body part |
| 11 | 997.0 | Nervous system complications |
| 12 | 997.1 | Cardiac complications |
| 13 | 997.2 | Peripheral vascular complications |
| 14 | 997.3 | Respiratory complications |
| 15 | 997.4 | Digestive system complications |
| 16 | 997.5 | Urinary complications |
| 17 | 997.6 | Amputation stump complication |
| 18 | 997.7 | Vascular complications of other vessels |
| 19 | 997.9 | Complications affecting other specified body systems, not elsewhere classified |
| 20 | 998.0 | Postoperative shock |
| 21 | 998.1 | Hemorrhage or hematoma or seroma complicating a procedure |
| 22 | 998.2 | Accidental puncture or laceration during a procedure |
| 23 | 998.3 | Disruption of wound |
| 24 | 9984 | Foreign body accidentally left during a procedure |
| 25 | 998.5 | Postoperative infection |
| 26 | 998.6 | Persistent postoperative fistula |
| 27 | 998.7 | Acute reaction to foreign substance accidentally left during a procedure |
| 28 | 998.8 | Other specified complications of procedures, not elsewhere classified |
| 29 | 998.9 | Unspecified complication of procedure, not elsewhere classified |
| **30** | 999.0 | Generalized vaccinia |
| **31** | 999.1 | Air embolism |
| **32** | 999.2 | Other vascular complications |
| **33** | 999.3 | Other infection |
| **34** | 999.4 | Anaphylactic shock due to serum |
| **35** | 999.5 | Other serum reaction |
| **36** | 999.6 | ABO incompatibility reaction |
| **37** | 999.7 | Rh incompatibility reaction |
| **38** | 999.8 | Other infusion and transfusion reaction |
| **39** | 999.9 | Other and unspecified complications of medical care, not elsewhere classified |